\documentclass[preprint,12pt,authoryear]{elsarticle}

\usepackage{amssymb}
\usepackage{amsmath}
\usepackage{graphicx}
\usepackage{epstopdf}
\usepackage{subfigure}

\journal{arXiv}

\begin{document}

\begin{frontmatter}



\title{Option Pricing with Convolutional Kolmogorov-Arnold Networks} 


\author{Zeyuan Li\textsuperscript{a},   Qingdao Huang\textsuperscript{a}} 

\affiliation{organization={School of Mathematics, Jilin University},
            addressline={2699 Qianjin Street, Changchun, Jilin Province}, 
            city={Changchun},
            postcode={130012},
            state={Jilin Province},
            country={China}}
\begin{abstract}
With the rapid advancement of neural networks, methods for option pricing have evolved significantly. 
This study employs the Black-Scholes-Merton (B-S-M) model, incorporating an additional variable to improve the accuracy of predictions compared to the traditional Black-Scholes (B-S) model. 
Furthermore, Convolutional Kolmogorov-Arnold Networks (Conv-KANs) and Kolmogorov-Arnold Networks (KANs) are introduced to demonstrate that networks with enhanced non-linear capabilities yield superior fitting performance. 
For comparative analysis, Conv-LSTM and LSTM models, which are widely used in time series forecasting, are also applied. 
Additionally, a novel data selection strategy is proposed to simulate a real trading environment, thereby enhancing the robustness of the model.
\end{abstract}


\begin{keyword}
Option pricing \sep Convolutional Kolmogorov-Arnold Networks \sep B-S-M model \sep Neural networks


\end{keyword}

\end{frontmatter}




\section{Introduction}
\label{sec1}
Due to the fact that even a small error can lead to significant losses, option pricing is always accompanied by high risk. 
Scholars from various fields have shown interest in option pricing, and as a result, many methods and theories have been developed. 
Among these methods, the most influential and practical guidance comes from the fields of theoretical finance and neural networks.

In theoretical finance, the Black-Scholes (B-S) model, introduced by Black and Scholes in 1973 [1], is the most renowned and widely applied model.
In the same year, Merton modified certain assumptions of the B-S model and proposed an improved version [2].
This enhanced model not only addresses the pricing of options in the presence of dividends but also notably influences subsequent research on implied volatility in financial markets.
This model is alternatively known as the Black-Scholes-Merton (B-S-M) model. However, in practical applications, both the B-S and B-S-M models exhibit biases and limitations due to the assumptions underlying these models, such as the No Arbitrage Principle and the allowance for continuous trading.

Hence, some experts began to focus on the emerging field of neural networks, combining them with traditional methods to conduct investigations, which yielded promising results.
In 1994, Hutchinson J.M. et al. pioneered the exploration of network models in the financial domain [10], incorporating various methods such as ordinary least squares, radial basis function, multi-layer perceptrons (MLPs), and projection pursuit.
Gençay R. et al. introduced techniques like hinting and Bayesian regularization, and elaborated on their roles [11][12].
Simultaneously, some scholars sought to apply genetic algorithms to the field of option pricing [13].
Nonetheless, the breakthrough in this field belongs to the development of recurrent neural networks (RNNs), because the data of options have a relatively strong temporal nature. One of the most famous of RNNs is the LSTM model, consequently, numerous investigators have concentrated on forecasting option price by leveraging LSTM [3][14].

A crucial innovation in neural networks is the development of KANs by Liu Z et al. [7].
The KANs have better interpretability for nonlinear fitting as well as accuracy, and in some other fields, scholars have carried out relevant simulations and obtained good results. 
Bodner A.D. and Drokin I. et al. have refined and developed the Conv-KANs model [8][9].
Nevertheless, there are no investigators who have applied these models to the field of option pricing with neural networks.

In this paper, the following contributions are presented. First, the B-S-M formula is utilized to achieve greater precision compared to the conventional B-S model.
Although this model has been established in theoretical finance for many years, it remains insufficiently exploited in the literature by scholars in this field.
Second,a novel data-processing method is also proposed to better approximate the real financial market environment, where the generalization capability of neural networks is enhanced by incorporating white noise into both the training and testing datasets. 
Finally, the popular KANs model and its convolutional version, Conv-KANs, are introduced, representing a new application of neural networks in the domain of option pricing. Experimental results show that these models outperform the original MLP-based models.

\section{Related Work}
\subsection{Kolmogorov-Arnold Networks}
\label{subsec1}
Before delving into KANs, it is essential to explore their mathematical theoretical foundations. 
Kolmogorov-Arnold representation theorem provides a promising platform to generate frameworks of neural networks, and many academics have studied that. 
Yet, this theorem has strict framework requirements.

The theorem was established by Arnold V I and  Kolmogorov A N. 
It is stated as suppose \textit{f} is a multivariate continuous function on a bounded domain, then \textit{f} can be written as a finite composition of continuous functions of a single variable and the binary operation of addition. 
More specifically, for a smooth $f : [0,1]^n \rightarrow \mathbb{R}$
\begin{equation}
  f(\mathbf{x})=f(x_1,\cdots,x_n)=\sum_{i=1}^{2n+1}\Phi_i\left(\sum_{p=1}^n\phi_{i,p}(x_p)\right)
 \end{equation}
 where

 \centerline{$\phi_{i,p}{:}[0,1]\to\mathbb{R}\mathrm{~and~}\Phi_i{:}\mathbb{R}\to\mathbb{R}$}

 On the one hand, this theorem seems to be very rigorously argued, but on the other hand, the framework for functions stuck with depth-2, width-(2n + 1) representation is too restrictive. 
 As a result, in the application of this construction, although there have been previous studies, the structure of the model is strictly limited to a relatively simple level, and it cannot compete with other MLPs with complex structures. 
 It is hoped that in paper [7], Liu Z et al. creatively propose the use of other frameworks to replace the original fixed framework and the results are promising. Here are their definitions of KAN:
\begin{equation}
  \mathbf{\Phi}=\{\phi_{i,p}\},p=1,2,\cdots,n_{\mathrm{in}},i=1,2\cdots,n_{\mathrm{out}}
\end{equation}

Where the parameters of the functions $\phi_{i,p}$ possess trainable properties. It supposes that every KAN layer which has $n_{\mathrm{in}}$-dimensional inputs and $n_{\mathrm{out}}$-dimensional outputs. And it is the 1D form.

When it refers to multi-dimensional situation, it tells
\begin{equation}
  \mathbf{x}_{l+1}=\underbrace{\begin{pmatrix}\phi_{l,1,1}(\cdot)&\phi_{l,1,2}(\cdot)&\cdots&\phi_{l,1,n_l}(\cdot)\\\phi_{l,2,1}(\cdot)&\phi_{l,2,2}(\cdot)&\cdots&\phi_{l,2,n_l}(\cdot)\\\vdots&\vdots&\cdots&\vdots\\\phi_{l,n_{l+1},1}(\cdot)&\phi_{l,n_{l+1},2}(\cdot)&\cdots&\phi_{l,n_{l+1},n_l}(\cdot)\end{pmatrix}}_{\mathbf{\Phi}_l}\mathbf{x}_l
\end{equation}

where ${\Phi}_{l}$ represents to the $l^{th} $ layer of the whole KAN framework while $x_{l} $ and $x_{l+1} $ separately denote input and output, respectively. 
Based on definitions (1) and (3), the structural formula for the complete KAN model can be derived directly. it is
\begin{equation}
  \mathrm{KAN}(\mathbf{x})=(\mathbf{\Phi}_{L-1}\circ\mathbf{\Phi}_{L-2}\circ\cdots\circ\mathbf{\Phi}_0)\mathbf{x}
\end{equation}
also
\begin{equation}
  f(\mathbf{x})=\sum_{i_{L-1}=1}^{n_{L-1}}\phi_{L-1,i_L,i_{L-1}}\left(\sum_{i_{L-2}=1}^{n_{L-2}}\cdots\left(\sum_{i_0=1}^{n_0}\phi_{0,i_1,i_0}(x_{i_0})\right)\cdots\right)
\end{equation}

Liu Z et al. demonstrate that this seemingly simple modification enables KANs to surpass MLPs in both accuracy and interpretability on small-scale AI + Science tasks [7]. 
See the paper for the proof and detailed reasoning process. 

\subsection{Convolutional Kolmogorov-Arnold Networks}
\label{subsec2}
In Convolutional Kolmogorov-Arnold Networks, The function $\phi$ is chosen as the kernel function, and in the original Kolmogorov-Arnold Networks, it serves as the activation function
\begin{equation}
  \phi=w_1\cdot spline(x)+w_2\cdot silu(x)
\end{equation}

In paper [8], Convolutional Kolmogorov-Arnold Networks were presented with 2-dimensions image $i$ as input. Suppose $\phi_{ij}$ to the corresponding pixel,
$a_{kl}$ and calculates the output pixel as the sum of $\phi_{ij}(a_{kl})$. Let $K$ be a KAN kernel $\in \mathbb{R}^{N\times M}$, it follows that
\begin{equation}
  (i*K)_{i,j}=\sum_{k=1}^N\sum_{l=1}^M\phi_{kl}(a_{i+k,j+l})
\end{equation}
and there is an example of kernel of Conv-KAN
\begin{equation}
  \text{KAN Kernel}=\begin{bmatrix}\phi_{11}&\phi_{12}&\phi_{13}\\\phi_{21}&\phi_{22}&\phi_{23}\\\phi_{31}&\phi_{32}&\phi_{33}\end{bmatrix}
\end{equation}
Equation (7) is a defined Conv-KAN. High dimensional version of Convolutional Kolmogorov-Arnold Networks, see paper [9].

\subsection{Conv-LSTM}
\label{subsec3}
The convolutional comparison experiment approach is transferred from KANs to LSTM, thereby introducing its classical variant, the Convolutional LSTM (Conv-LSTM).
Naturally, just like the LSTM model, it excels in dealing with the prediction of time series. 
It was developed to address the problem of precipitation forecasting, which involves predicting the precipitation for the following $H$ hours based on the previous $B$ observations [6]:
\begin{equation}
  \tilde{X}_{t+1},\cdots,\tilde{X}_{t+H}=\arg\max_{X_{t+1},\cdots,X_{t+H}}p(X_{t+1},\cdots,X_{t+H}|\hat{X}_{t-B+1},\cdots,\hat{X}_{t})
\end{equation}

The contribution of Conv-LSTM lies in combining the convolution operation, which extracts spatial features, with the LSTM, which captures temporal features [3].

In Conv-LSTM, the input, hidden state, and memory cell are represented as matrices, which distinguishes it from the classical LSTM, 
 where  ‘$\circ$’ denotes the Hadamard product and ‘$\ast$’ represents the convolution operator:
\begin{equation}
  \begin{gathered}
i_t=\sigma(W_{xi}*X_t+W_{hi}*H_{t-1}+W_{ci}\circ C_{t-1}+b_i) \\
f_t=\sigma\big(W_{xf}*X_t+W_{hf}*H_{t-1}+W_{cf}\circ C_{t-1}+b_f\big) \\
C_t=f_t\circ C_{t-1}+i_t\circ\tanh(W_{xc}*X_t+W_{hc}*H_{t-1}+b_c) \\
o_t=\sigma(W_{xo}*X_t+W_{ho}*H_{t-1}+W_{co}\circ C_t+b_o) \\
H_{t}=o_{t}\circ\mathrm{tanh}(C_{t}) 
\end{gathered}
\end{equation}

In paper [3], Ge M., Zhou S., Luo S., et al., propose 1D and 3D convolution variants of Conv-LSTM and conduct investigations, aiming to achieve improved training results. 
This is important work, alternatively, for comparison purposes, the basic 1D version is chosen. 

\section{Model Structure}
In this paper, three parts are employed to conduct the experiment. Part I discusses classical financial theory formulas. 
While formula (13) is commonly used by other researchers, an improvement has been made, and formula (14) is concurrently applied for comparison. 
Part II discusses the application of the Conv-LSTM and LSTM models, which are widely used in capturing features of time series data. 
Part III introduces the new models, KANs and Conv-KANs. Traditional KANs have stronger adaptability and accuracy for fitting nonlinear functions, they are also a promising alternative to the classic MLPs model.
\subsection{B-S-M Formula}
The Black–Scholes–Merton differential equation governs the pricing of any derivative based on a non-dividend-paying stock [1][2].
The proof for this part is provided in the appendix.

Assume that $f$ represents the price of a call option or other derivatives dependent on $S$. While $T$ is the maturity date, $t$ is the general time at which the price of derivatives is considered, 
and $r$ is the associated risk-free interest rate. It follows that

\begin{equation}
  \frac{\partial f}{\partial t}+rS\frac{\partial f}{\partial S}+\frac{1}{2}\sigma^{2}S^{2}\frac{\partial^{2}f}{\partial S^{2}}=rf
\end{equation}

Equation (11) represents the Black–Scholes–Merton differential equation. It admits various solutions corresponding to different derivatives defined in terms of $S$. 
The variable solution of (11) can potentially be determined based on the specific boundary conditions. The boundary conditions of options were addressed in the introduction section. When it comes to call option, it is
\begin{equation}
  f=max(S-K,0)\quad\mathrm{when}\quad t=T
\end{equation}
Accordingly, the boundary conditions of put option is

\centerline{$f=max(K-S,0)\quad\mathrm{when}\quad t=T$}

The most well-known solutions of equation (11) are the prices of call and put options, and the corresponding formulas are as follows
\begin{equation}
  \begin{gathered}
  c=S_0N(d_1)-Ke^{-rT}N(d_2) \\ 
  p=Ke^{-rT}N(-d_2)-S_0N(-d_1)
  \end{gathered}
\end{equation}
Where
\centerline{$\begin{aligned}&d_{1}=\frac{\ln{(S_0/K)}+(r+\sigma^2/2)T}{\sigma\sqrt{T}}\\&d_{2}=\frac{\ln{(S_0/K)}+(r-\sigma^2/2)T}{\sigma\sqrt{T}}=d_1-\sigma\sqrt{T}\end{aligned}$}
These formulas are widely used by contributors in the field of option pricing with neural networks to compare their frameworks, as demonstrated in paper [3]. 
It is observed that many options in financial markets are subject to dividends, due to the dividends paid by the underlying assets (stocks). 
Consequently, Merton's research [2] are utilized to replace the original no-dividends formula (13). 
A new variable $q$ is introduced, denoting the rate of dividend interest. They are
\begin{equation}
  \begin{gathered}
  c=S_{0}e^{-qT}N(d_{1})-Ke^{-rT}N(d_{2})\\p=Ke^{-rT}N(-d_{2})-S_{0}e^{-qT}N(-d_{1})
\end{gathered}
\end{equation}
Where

\centerline{$\ln\frac{S_0e^{-qT}}K=\ln\frac{S_0}K-qT$}

it follows that $d_1$ and $d_2$ are expressed as follows

\centerline{$\begin{aligned}&d_{1}=\frac{\ln{(S_0/K)}+(r-q+\sigma^2/2)T}{\sigma\sqrt{T}}\\&d_{2}=\frac{\ln{(S_0/K)}+(r-q-\sigma^2/2)T}{\sigma\sqrt{T}}=d_1-\sigma\sqrt{T}\end{aligned}$}

Experimental results indicate that this formula provides greater accuracy compared to the classical B-S formula. From the perspective of theoretical finance, the B-S-M model becomes more accurate with the incorporation of the variable 
$q$, a conclusion further validated by the experimental results in this paper. From the perspective of network learning, the B-S-M model has guided us in identifying and introducing the new variable 
$q$. This variable had long been overlooked in prior studies on neural network-based option pricing, particularly because some stocks pay dividends.

\subsection{Conv-LSTM}
In this paper, a 1D Conv-LSTM is selected for comparison with Conv-KANs. However, in the data processing process, 
a three-dimensional processing approach was used, and the data were later converted into a one-dimensional form. 
This means that the 1D input form was ultimately chosen, and the data are arranged as shown in Figure 1.
\begin{figure}[htbp]
  \centering
  \includegraphics[width=0.8\linewidth]{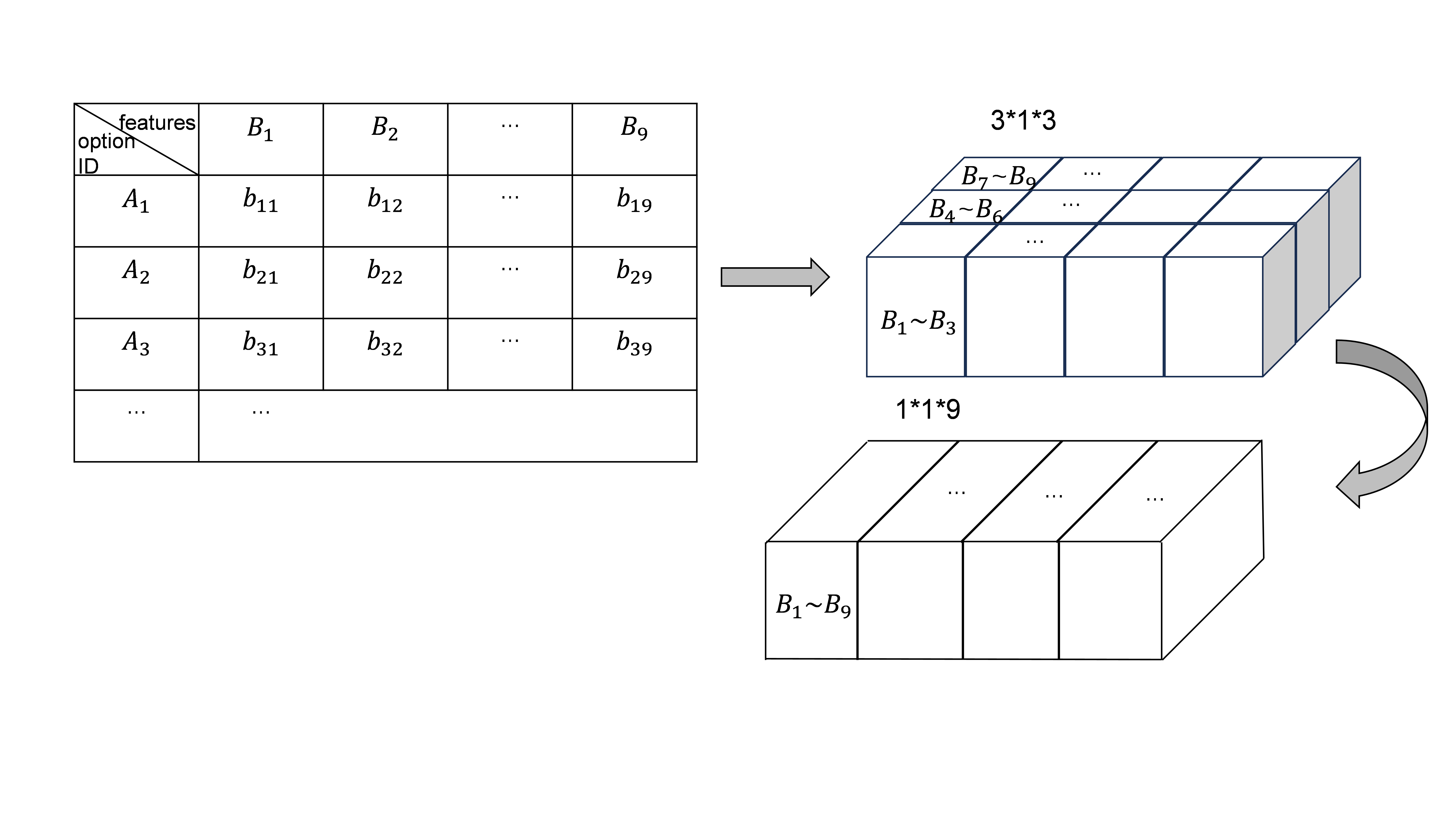}
  \caption{Feature representation}
  \label{fig1.}
\end{figure}

From figure1, it can be concluded that the input for a single day's option is ($C,N,D$),
where $C=1$ and $N$, representing the numbers of channels and observations, respectively,  while $D=9$ denotes the number of data features.

In the original paper [6], Conv-LSTM is intended to capture spatiotemporal correlations. 
The entire dataset for input has the shape of ($T,C,N,D,E$), $T$ is the whole quantities of dataset, $E$ is the second dimension of the image data. 
Despite this, the format of the option data is shown in the Figure 1 above, and clearly the data have only one dimension. 
Taking these factors into account, the input used in the experiment is ($T,C,N,D$). The values for $C$ and $D$ have been mentioned above.

\begin{figure}[htbp]
  \centering
  \includegraphics[width=0.8\linewidth]{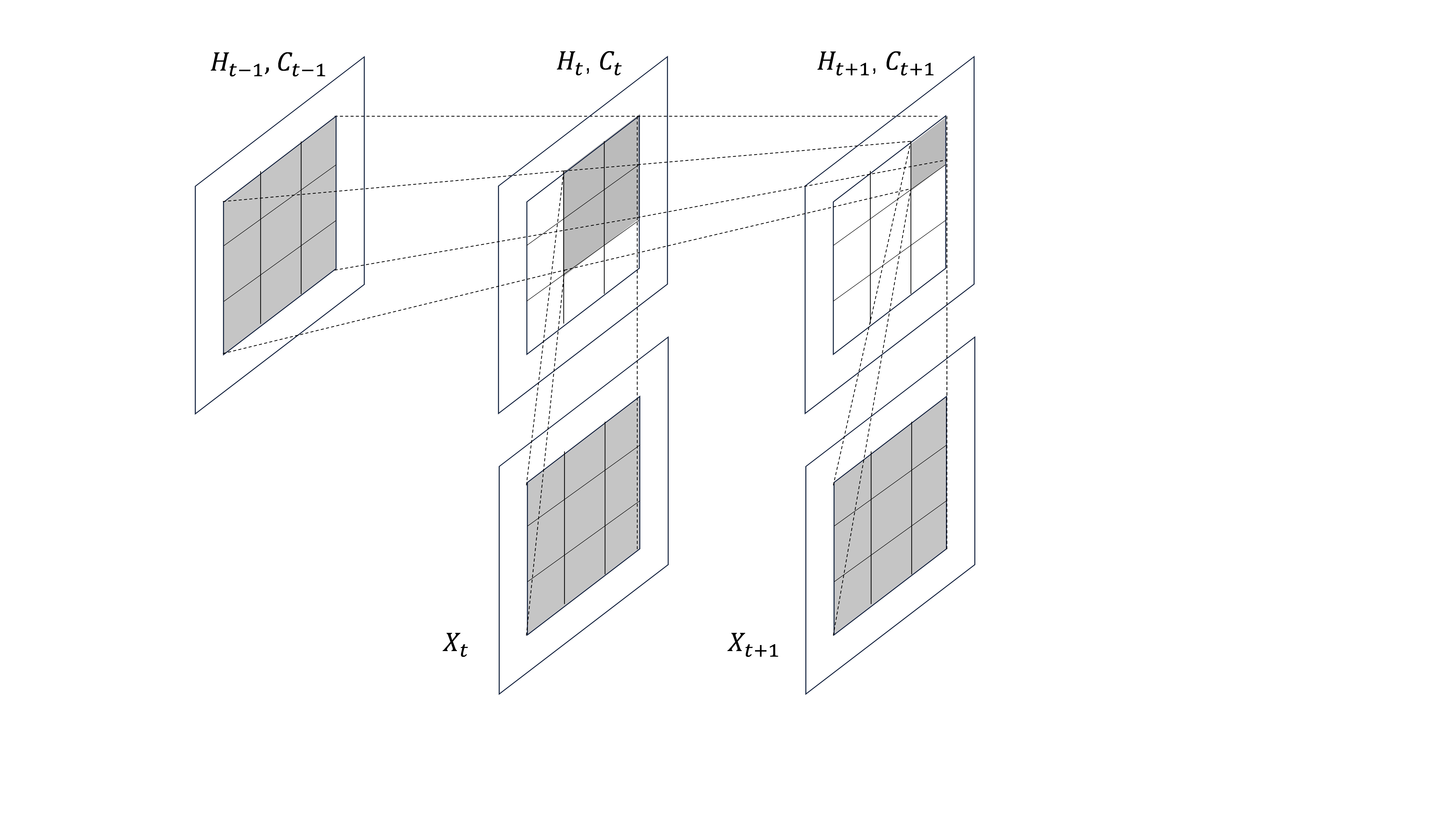}
  \caption{The architecture of the Conv-LSTM model
  }
\end{figure}

\subsection{KANs}
In this study, the input data is consistently in the form of ($T,C,N,D$). But in the primary work of Liu Z et al., the focus is on simple functions (usually two dependent variables). 
Thus, the initial model code can only hold a maximum of three input variables, such as ($C,N,D$). The code is improved to allow the input of four or more variables.

From the Related Work Chapter, it can be concluded that the non-linear nature of KANs arises from equation (6). 
In contrast to traditional MLPs, where nonlinear behavior arises from the linear combination of inputs followed by the application of an activation function, 
the nonlinear behavior in this model is directly achieved through the combination of nonlinear functions. This can also be understood as the combination of all activation functions.

\begin{figure}[htbp]
  \centering
  \includegraphics[width=0.8\linewidth]{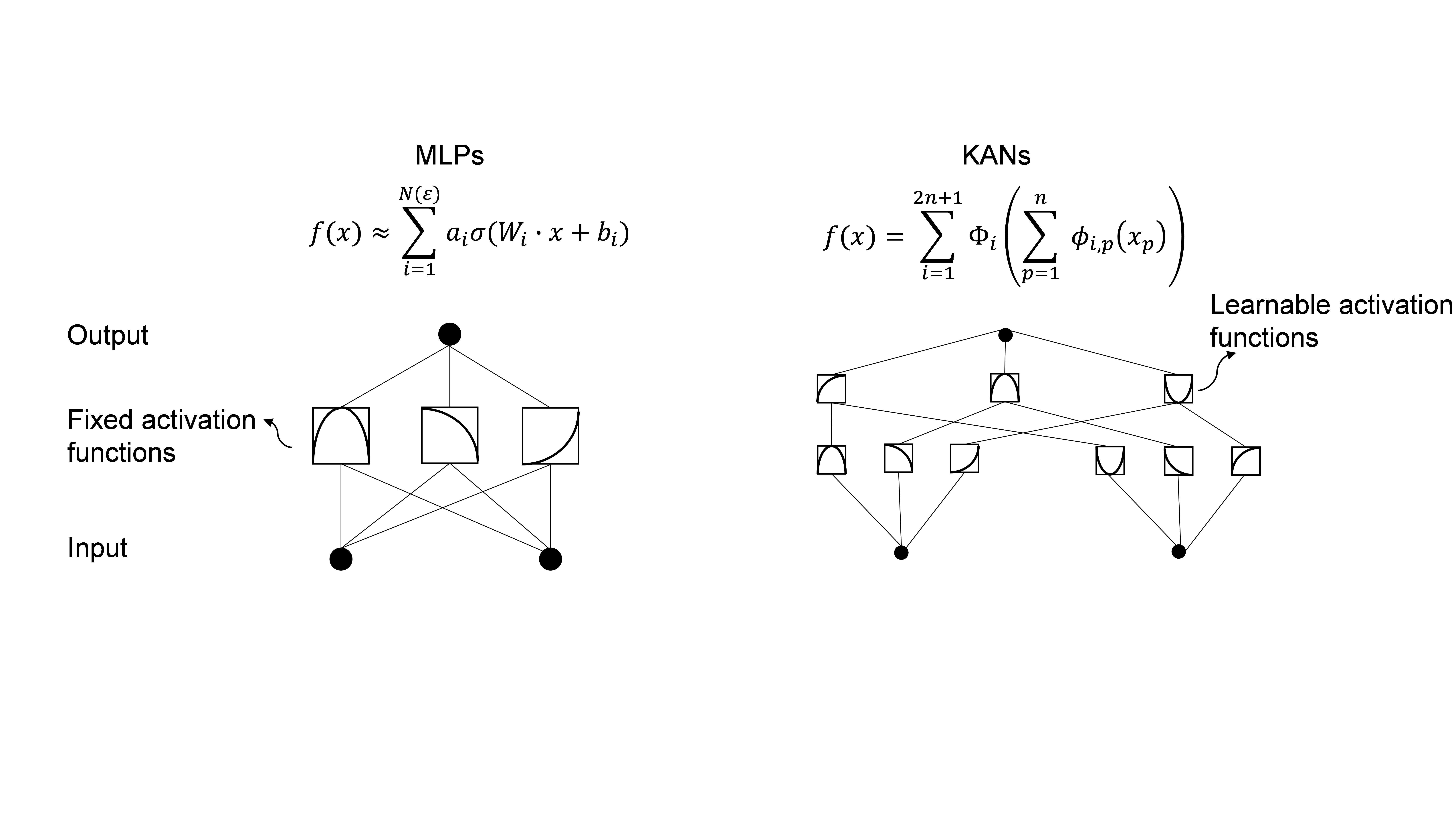}
  \caption{The structures of MLPs and KANs}
\end{figure}

In this paper, our aim is to capture the nonlinear nature of the formula in option pricing, and Nonlinear formulas from theoretical finance, such as the B-S-M formula (14), are listed, 
which also hold significant guiding value for pricing in real markets. Liu Z et al. noted that, 
for accuracy, smaller KANs can achieve comparable or even better performance than larger MLPs in function fitting tasks [7]. And this conclusion is also proved by our investigation 

\subsection{Conv-KANs}
In the context of Conv-KANs, it is essential to review the convolutional variant of the MLPs model.

\begin{figure}[htbp]
  \centering
  \includegraphics[width=0.8\linewidth]{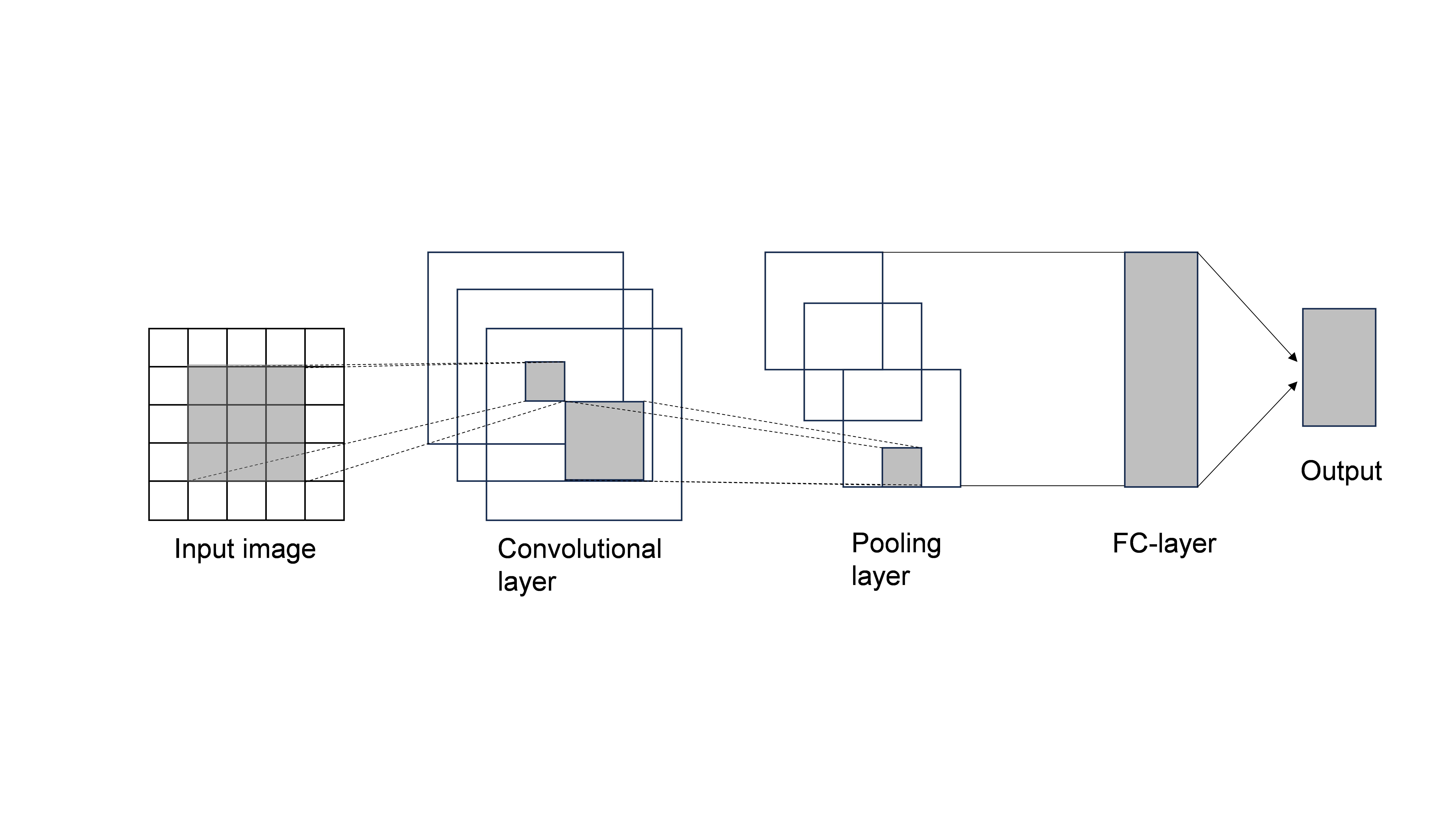}
  \caption{An example of convolutional neural network}
\end{figure}

In fact, in convolutional neural networks (CNNs), the indispensable parts include input, convolutional layer, fully connected layer, and output. 
In the preceding section, it was established that one of the key distinctions between KANs and MLPs lies in the method by which the nonlinear property is derived. 
When it comes to convolution operations, the first issue to address is whether the fully connected layer should be modified, as the fully connected layer has long been associated with MLPs. 
This leads to the question of whether, when building Conv-KANs, the entire architecture should be based on convolutions as in KANs, or if MLPs should be used for the fully connected layer. 

\begin{figure}[htbp]
  \centering
  \includegraphics[width=0.8\linewidth]{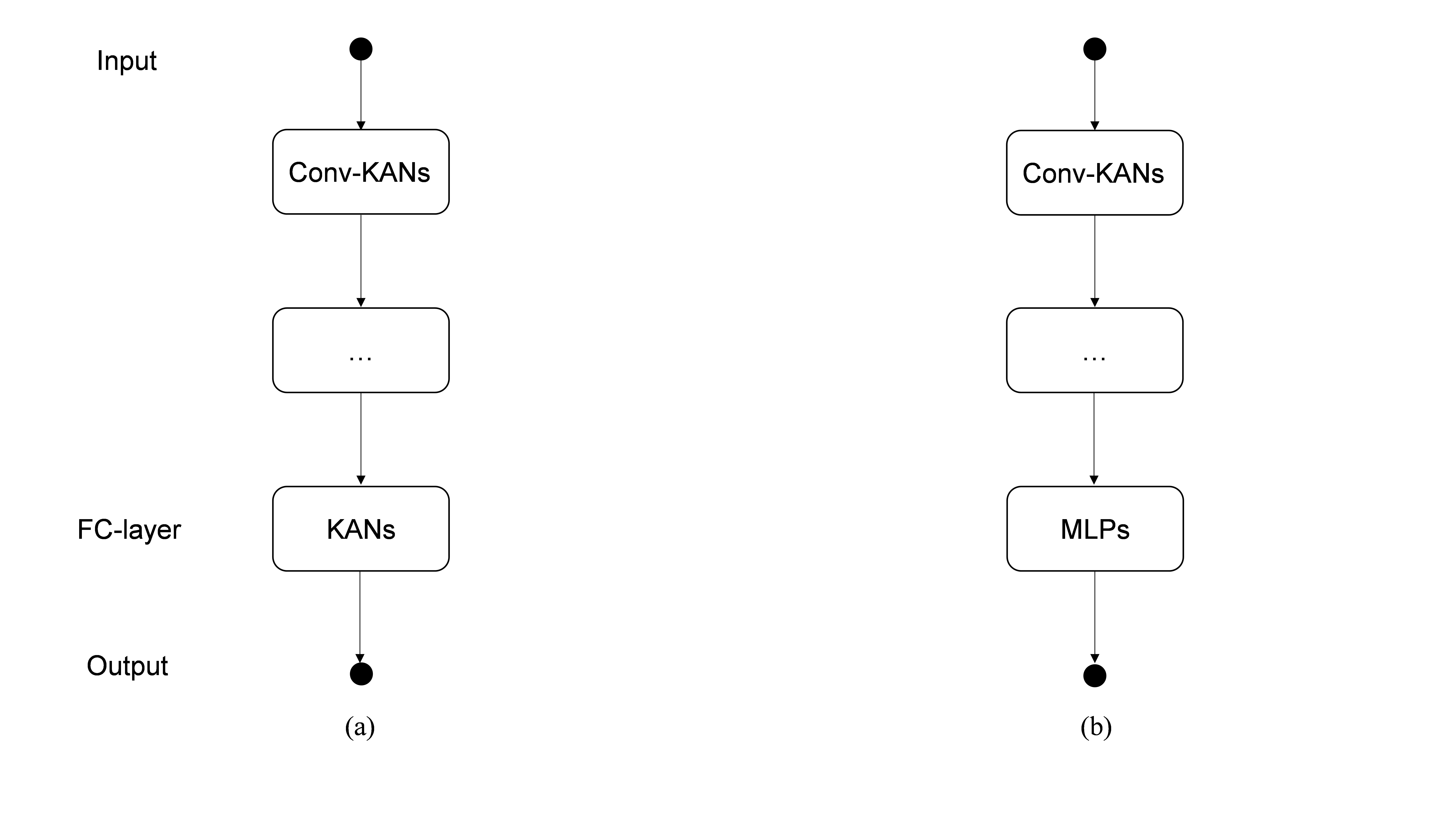}
  \caption{Which one to choose}
\end{figure}

Ultimately, the choice is made to retain the traditional fully connected layer, on the one hand, to facilitate comparison with Conv-LSTM. 
On the other hand, the study of model architecture that was built entirely by KANs was insufficient. The use of the fully connected layer may better reflect the difference between the two models.
In the experiment, these architectures are used as shown in Figure 6.

\begin{figure}[htbp]
  \centering
  \includegraphics[width=0.8\linewidth]{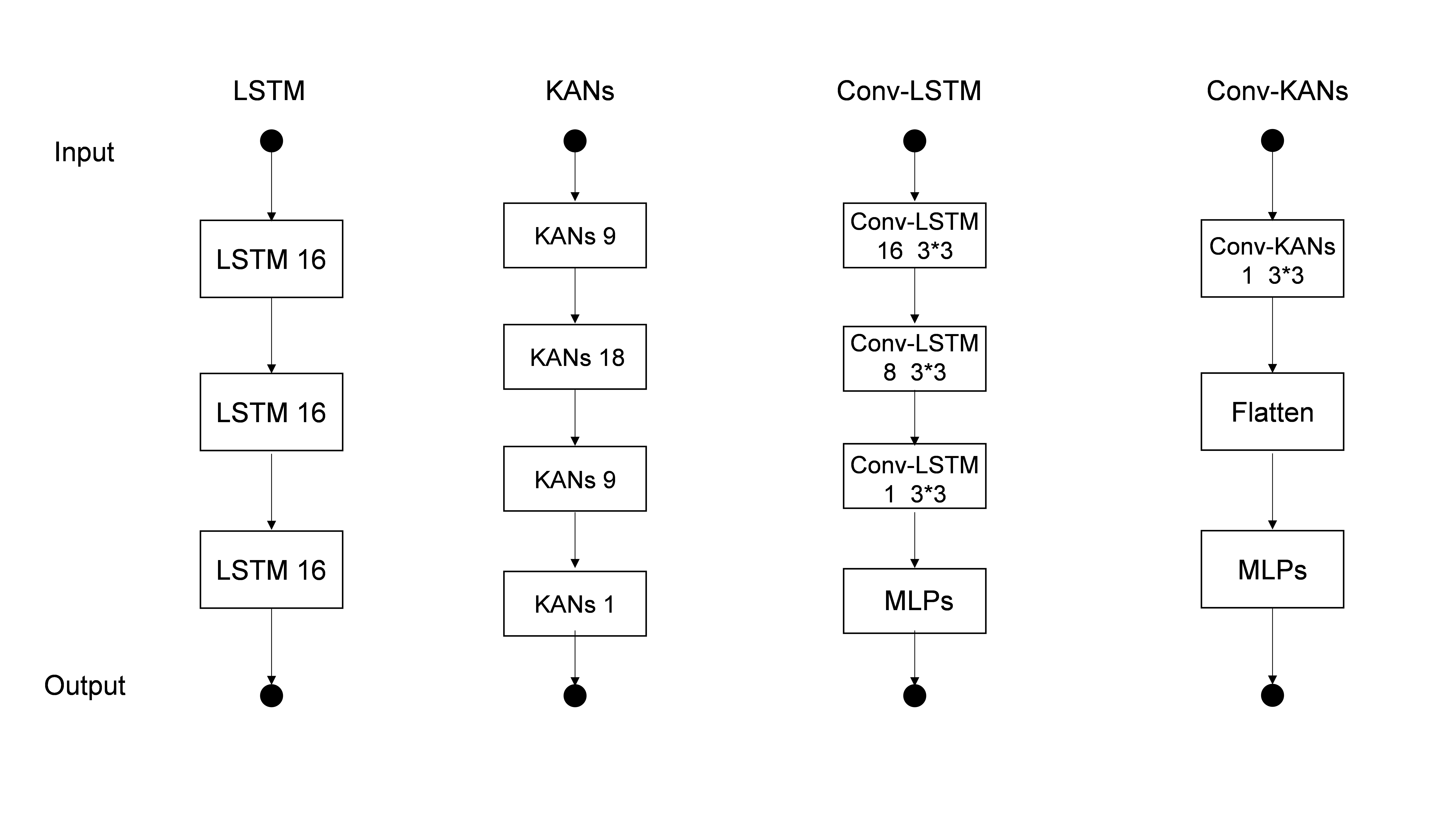}
  \caption{Architectures used in the experiments}
\end{figure}

\section{Experiment}
\subsection{Data and Data Processing Method}
In our experiment, CSI 300 Index option data collected from Chinese financial markets from January 1st, 2020, to December 31st, 2020, are utilized. 
The first eight months of the year are engaged as the training dataset, while the last four months serve as the test dataset. 
Nine variables were deployed in the study, namely time to maturity, option type, Delta, strike price, spot price, theoretical price, monthly dividend rate, 
risk-free rate, and volatility calculated using the GARCH model.
The training data consist of 42,125 observations, while the test data include 21,263 observations. The data used in this analysis are obtained from www.resset.com.

It is crucial to emphasize that, to better simulate a real trading environment, all actual data (except for a small amount of missing data) are utilized, 
in contrast to other experts in the field who filter the data to remove noise based on moneyness ($S/K$). 
Investigations conducted by Ivașcu C F confirm that Machine Learning models outperformed by a great margin the parametric models [4]. 
Nevertheless, from the generalization point of view, too much data processing will affect the generalization ability of the network model [5]. 
And although many scholars tend to process data, they also apply similar processing to the test set data.
Therefore, the control of the real environment is weakened and the problems caused by the reduced generalization ability of the model are avoided. 

\begin{figure}[htbp]
  \centering
  \includegraphics[width=0.8\linewidth]{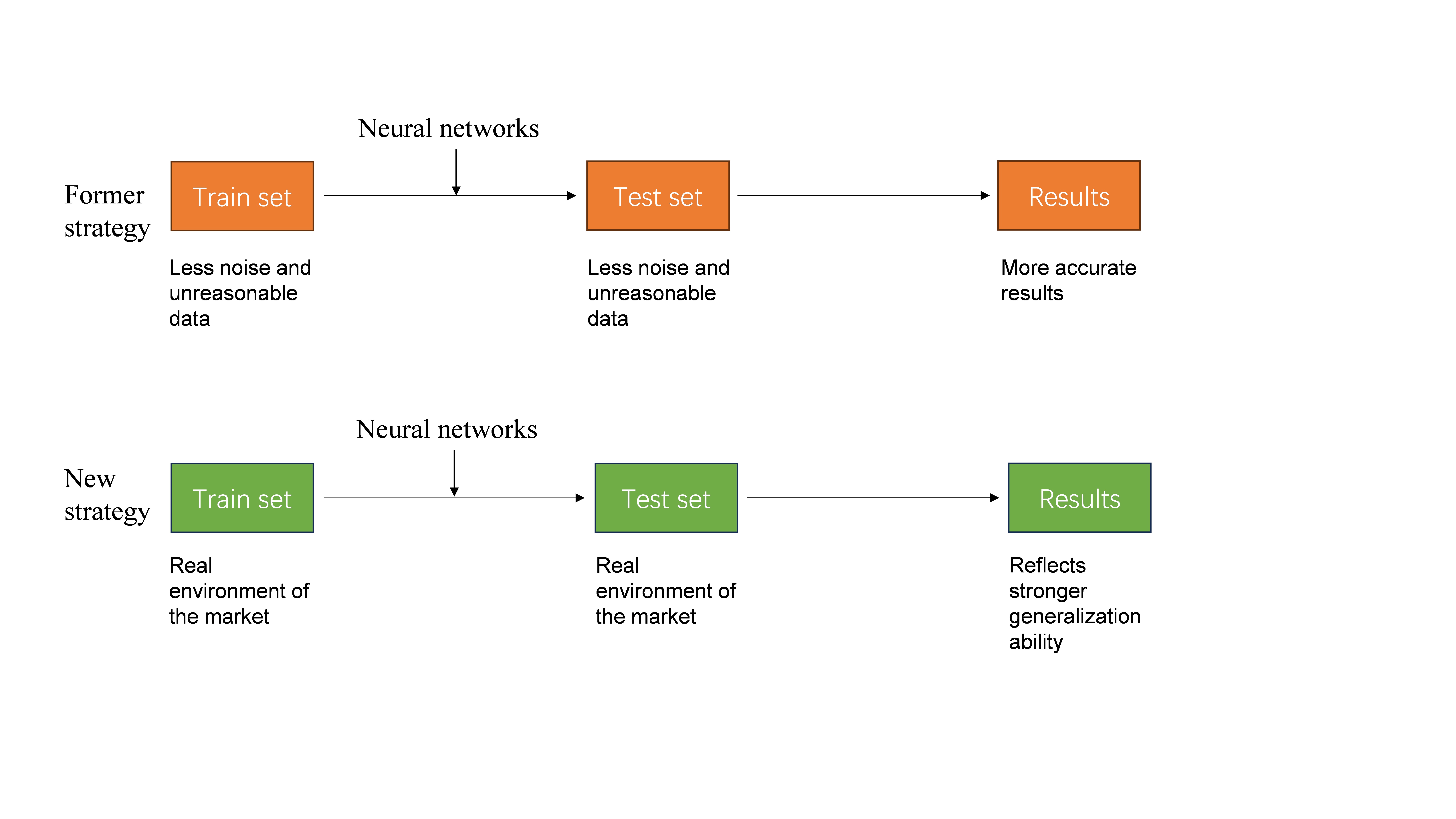}
  \caption{The results of the strategies}
\end{figure}

To facilitate comparison, a certain amount of noise is intentionally retained in the original data, 
ensuring that the trained model exhibits stronger generalization capabilities. Assuming that an option was issued on July 1st and exercised on September 15th, 
the investigators would normally include all the data from the time the option was sold to the time it was exercised in the training set. 
In contrast, the option data are divided into two segments: the information from July 1st to August 31st is allocated to the training set, and the remaining data are assigned to the test set.

\begin{figure}[htbp]
  \centering
  \includegraphics[width=0.8\linewidth]{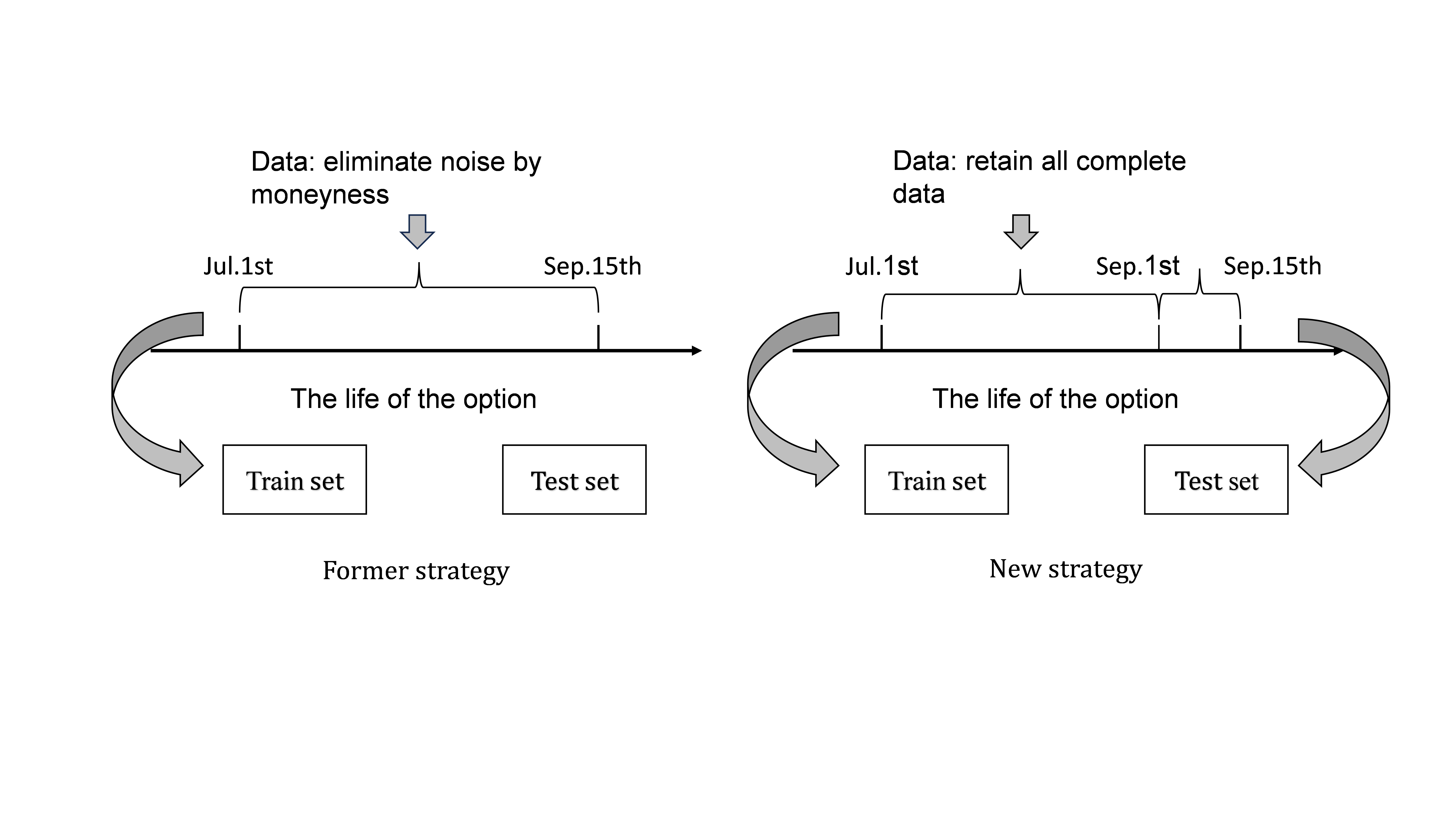}
  \caption{Strategies presentation}
\end{figure}

This leads to the extreme situation where the option offered on August 31st becomes a point of only one day of data in the training set, that is, white noise. 
The effects of increased noise and direct use of raw data are to sacrifice the accuracy of the model and improve generalization. 
The reference parametric model, such as B-S-M model, is unaffected by noise and time series, and thus irrefutably has relatively better results in this study. 
It is pleasing to note that, although the standard network model does not perform well when compared to the B-S-M model, its convolutional version outperforms the parametric model.

The option data are recorded sequentially in chronological order to align with the data characteristics necessary for applying LSTM and Conv-LSTM models.

In case the model tends to be imbalanced and asymmetric, each variable in the dataset must be normalized prior to the procedure. Specifically:

\begin{equation}
  \tilde{x}\quad=\quad\frac{x-\mu}{\sigma}
\end{equation}
and

\centerline{$\mu=\frac{1}{n}\sum_{i=1}^{N} x_{i},\sigma^{2}=\frac{1}{n}\sum_{i=1}^{N} (x_{i}-\mu)^{2}$}

\subsection{Empirical analysis}
The analysis is performed within the Pytorch environment, utilizing the B-S model, the B-S-M model, and the models illustrated in Figure 6. 
The mean squared error is adopted as the loss function, and the Adam optimizer is used for learning. 
KANs and Conv-KANs are trained for 50 epochs, whereas Conv-LSTM and LSTM are each trained for 200 epochs.
The batch size and the learning rate are selected to 32 and 0.00001.

The results of the experiment are shown in the Figures 9–14. For simplicity, only the options for the first 240 days are recorded.
 
\begin{figure}[htbp]
  \centering
  \includegraphics[width=0.8\linewidth]{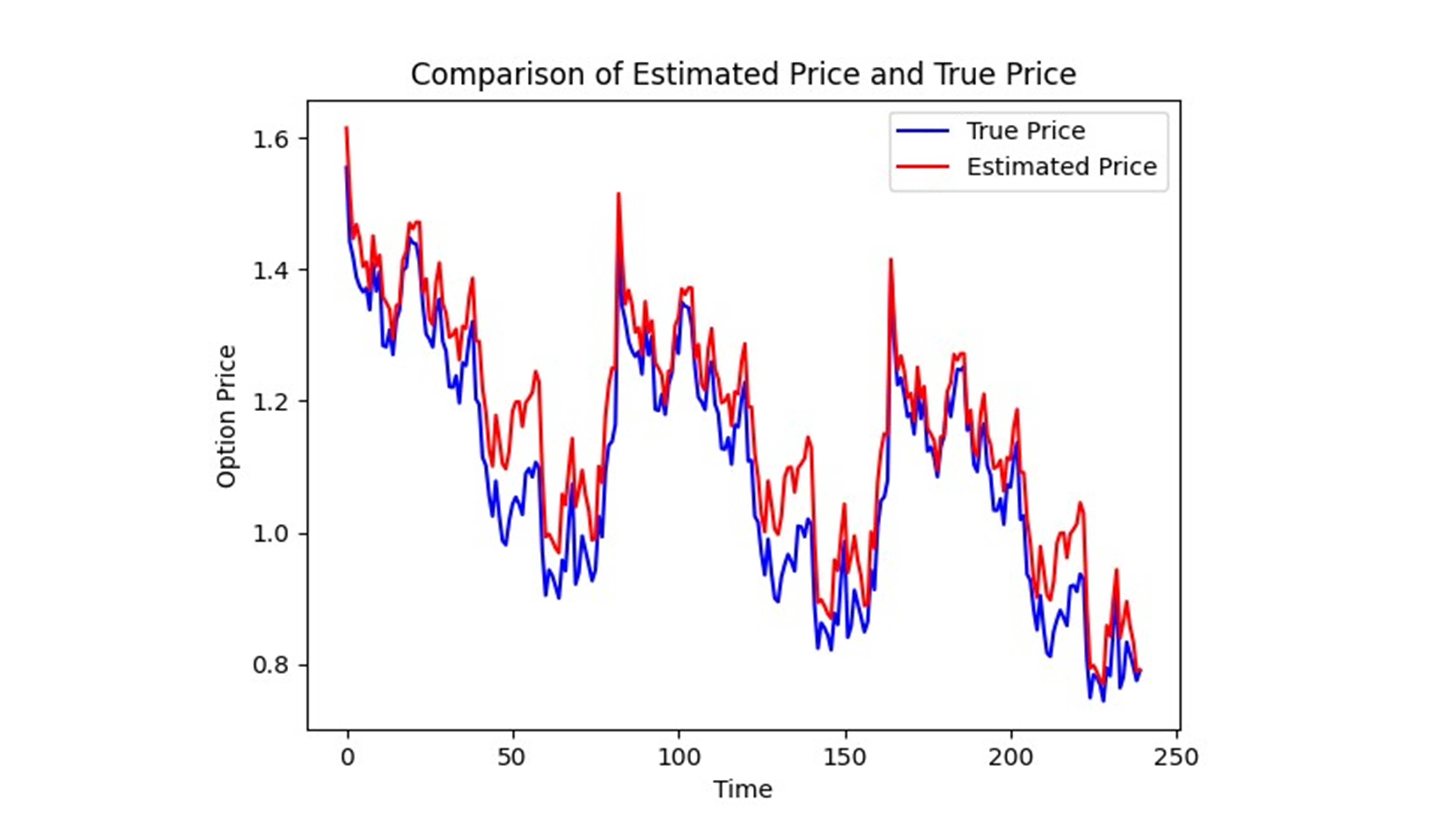}
  \caption{Forecasting result of B-S model}
\end{figure}

\begin{figure}[htbp]
  \centering
  \includegraphics[width=0.8\linewidth]{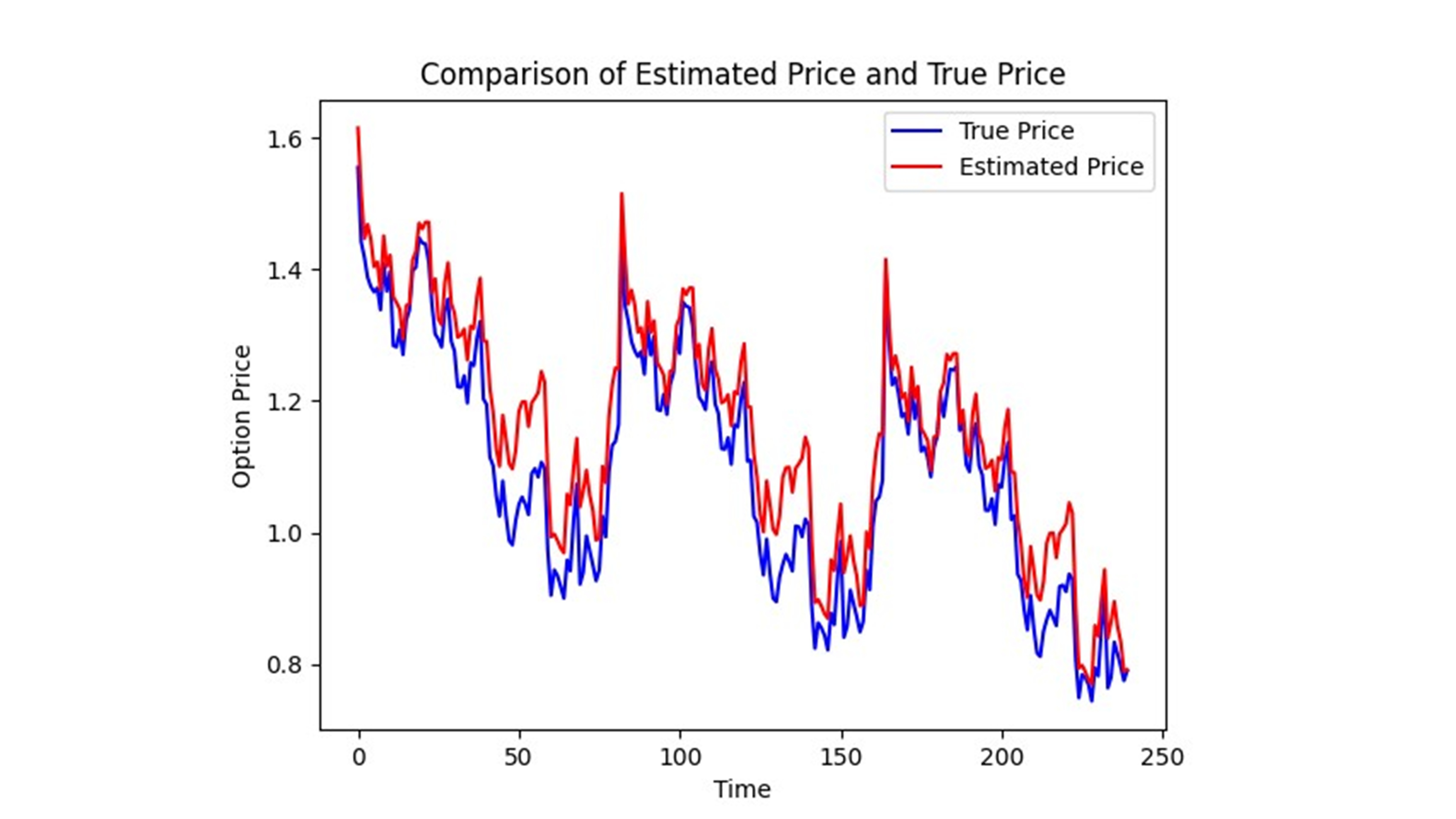}
  \caption{Forecasting result of B-S-M model}
\end{figure}

\begin{figure}[htbp]
  \centering
  \includegraphics[width=0.8\linewidth]{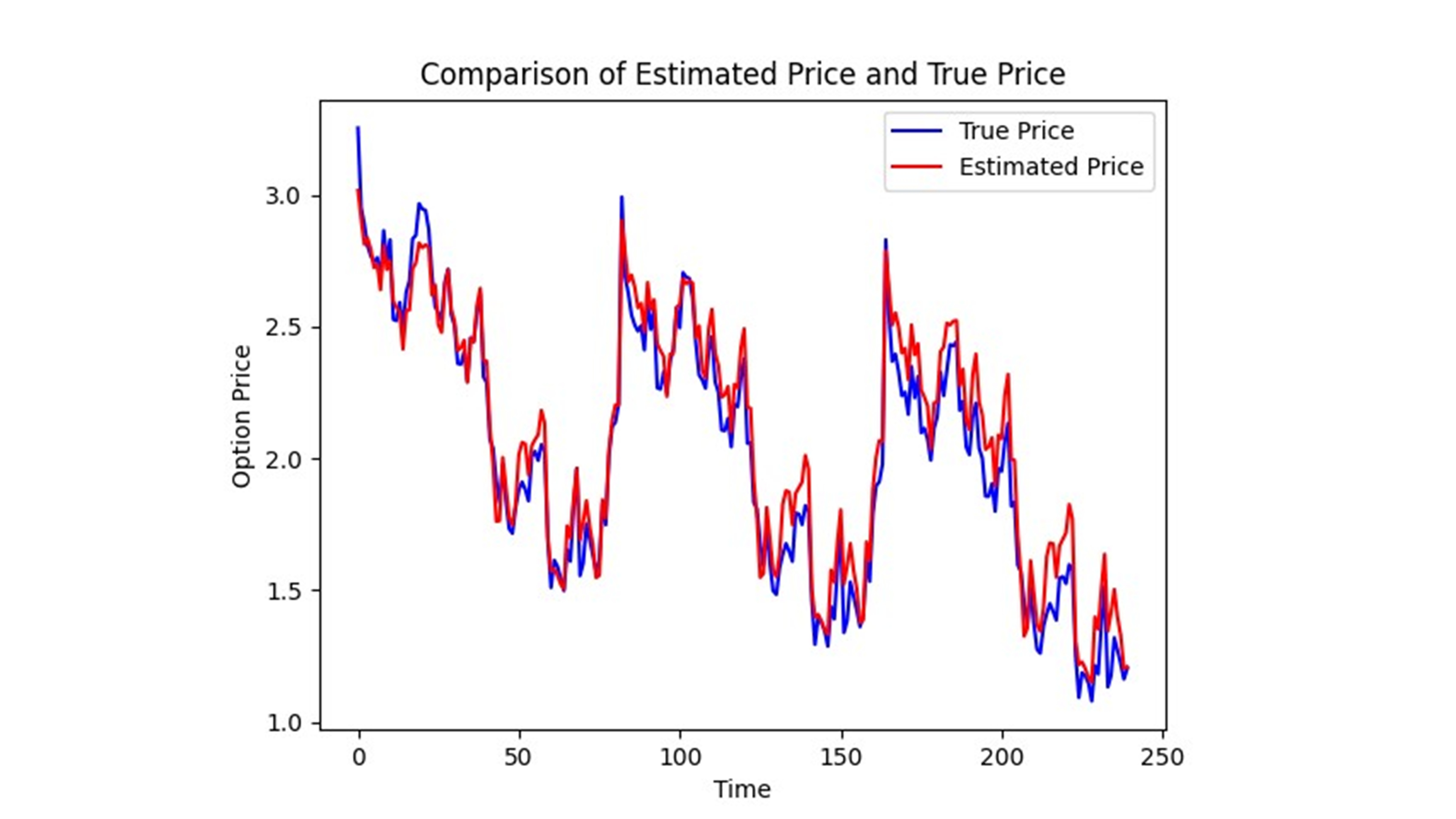}
  \caption{Forecasting result of LSTM model}
\end{figure}

\begin{figure}[htbp]
  \centering
  \includegraphics[width=0.8\linewidth]{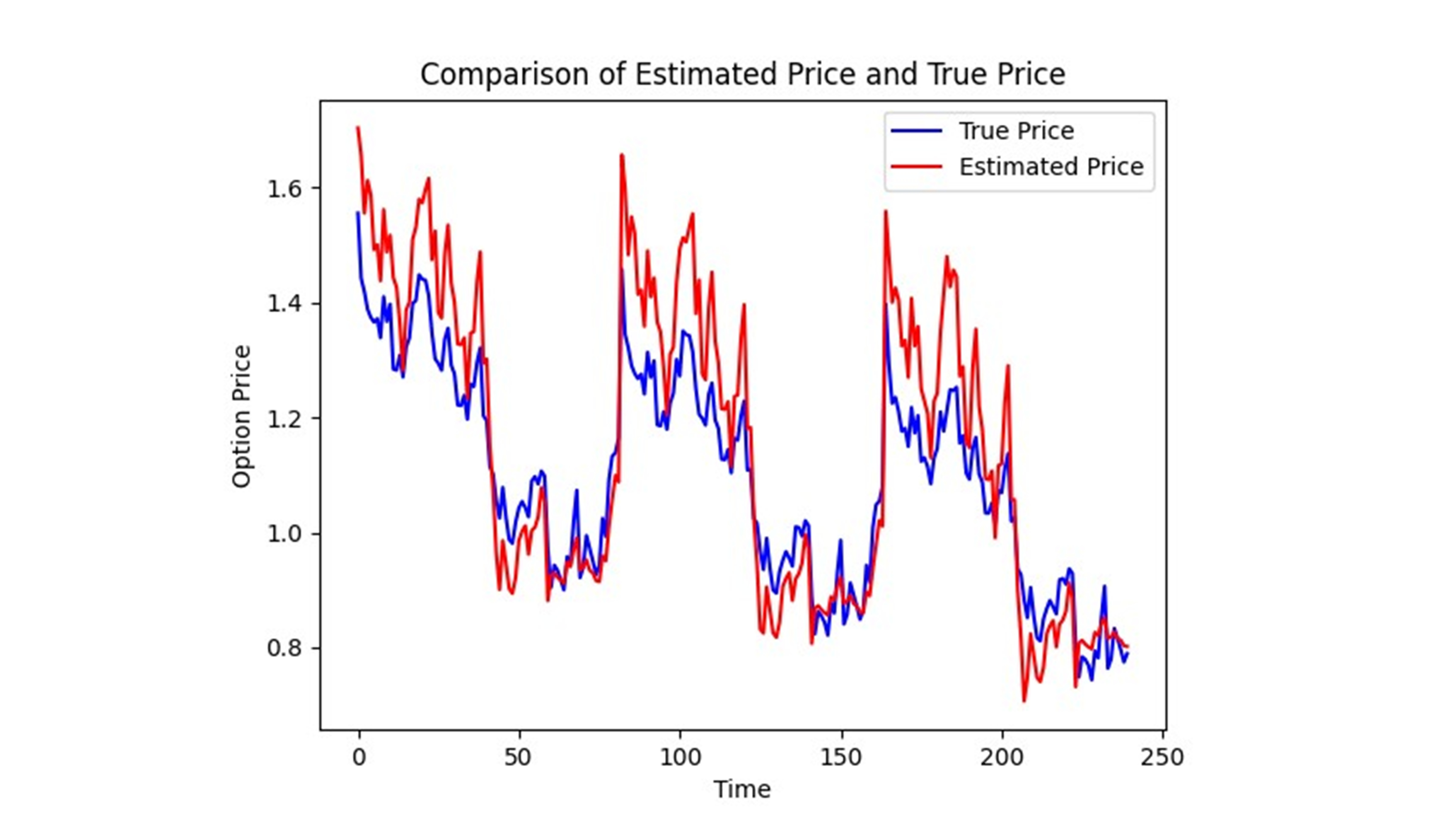}
  \caption{Forecasting result of Conv-LSTM model}
\end{figure}

\begin{figure}[htbp]
  \centering
  \includegraphics[width=0.8\linewidth]{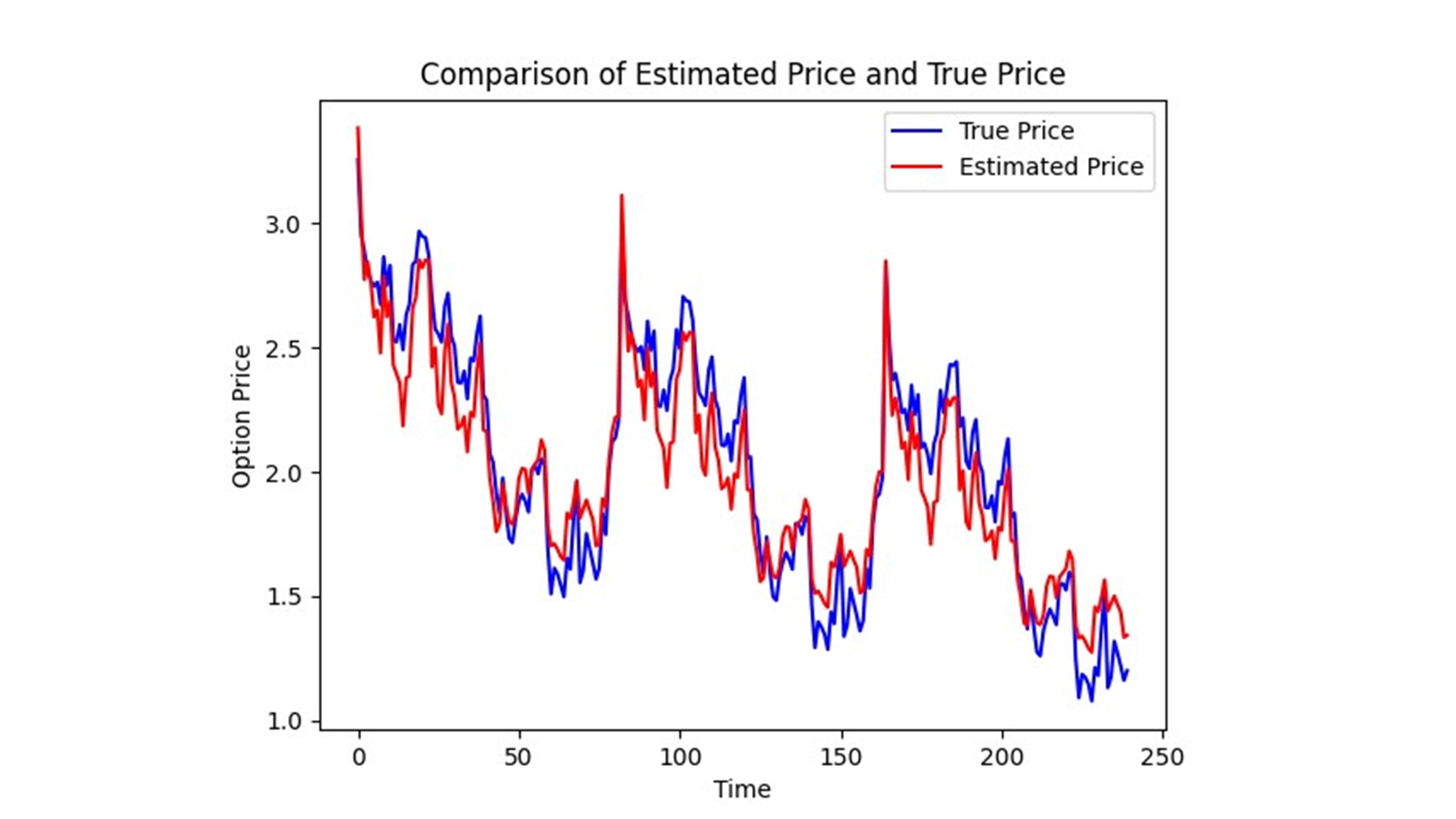}
  \caption{Forecasting result of KANs model}
\end{figure}

\begin{figure}[htbp]
  \centering
  \includegraphics[width=0.8\linewidth]{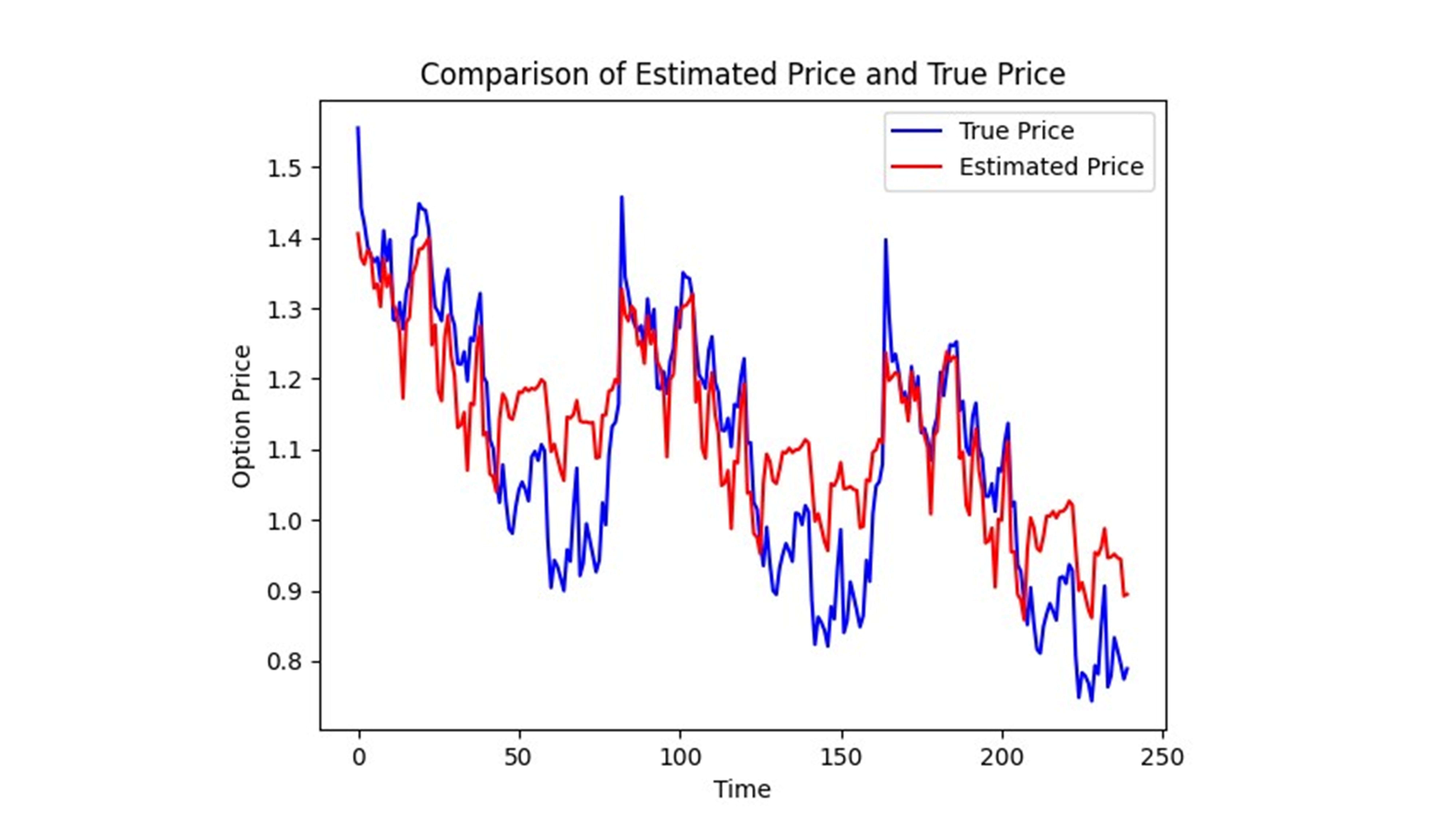}
  \caption{Forecasting result of Conv-KANs model}
\end{figure}

The results shown by the pictures are similar to other studies [3] in that the two lines of the figures from the B-S model match very well. 
The difference is that in this study, the correspondence of the two curves of the B-S-M model and LSTM is also excellent. 
Apart from these, the results of KANs are close to matching, but there are some slight differences compared to the former. 
Conv-KANs and Conv-LSTM show contrasting behavior, with one tending to forecast overall leveling off and the other predicting prices with larger peaks at the real price peaks.

To provide a more comprehensive evaluation, four performance metrics are selected: mean squared error (MSE), 
root mean squared error (RMSE), mean absolute error (MAE), and mean absolute percentage error (MAPE). The following are the calculation formulas:

\begin{equation}
  \begin{gathered}
  \mathrm{MSE}=\frac{1}{N}\sum_{i=1}^{N} (y_{i}-\hat{y}_{i})^{2}\quad\mathrm{RMSE}=\sqrt{\frac{1}{N}\sum_{i=1}^{N} (y_{i}-\hat{y}_{i})^{2}}\\
  \mathrm{MAP}=\frac{1}{N}\sum_{i=1}^{N} |y_{i}-\hat{y}_{i}|\quad\mathrm{MAPE}=\frac{1}{N}\sum_{i=1}^{N} |\frac{y_{i}-\hat{y}_{i}}{y_{i}}|^{\leftarrow}
  \end{gathered}
\end{equation}

Table 1 shows the results:
\begin{table}[h!]
  \begin{center}    
    \begin{tabular}{c c c c c c} 
      \hline
      \textbf{   } & \textbf{MSE} & \textbf{RMSE}& \textbf{MAP} & \textbf{MAPE}\\
      \hline
      B-S & 0.01612 & 0.12695 &0.08449&0.60950\\
      B-S-M & 0.01552 & 0.12457 &0.08442&0.61361\\
      LSTM & 0.03756 & 0.19380 &0.14574&0.78580\\
      Conv-LSTM & 0.01350 & 0.11620 &0.08808&12.36620\\
      KANs & 0.02679 & 0.16369 &0.12627&0.67222\\
      Conv-KANs & 0.00790 & 0.08890 &0.07098&15.11880\\
      \hline
    \end{tabular}
    \caption{The numerical error results of the models}
  \end{center}
\end{table}

The experimental results indicate that the data processing method adopted generates significant challenges for the simulation.
The performance of LSTM and KANs is worse than that of the B-S-M model, 
nevertheless, despite fewer training epochs and a simpler model architecture compared to LSTM, KANs yield relatively good results.
Meanwhile, CNNs have also shown very good results. In addition, three measurements by Conv-KANs are better than Conv-LSTM, 
however, in terms of MAPE, both results are worse than those of the other models, which may be due to the nature of the convolutional structures.

\section{Conclusion}
In today's complex financial derivatives pricing landscape, many researchers have turned to option pricing with neural networks, utilizing MLP-based models to achieve promising results.
As a promising alternative to MLP models, the KANs model has demonstrated strong performance in nonlinear fitting.
And this fits in well with nonlinear models in traditional financial theory. In this paper, the B-S model is improved for comparison, and the powerful LSTM model for time series data, along with its convolutional version, is introduced. 
For the first time, the KANs model and its convolutional version are employed, excelling at nonlinear fitting.
A new data processing method is also proposed, aimed at better simulating the realities of the financial market.

In the experiment, KANs once again demonstrated its strength in nonlinear fitting, and we also observed the excellent performance of CNNs, such as Conv-KANs, in option pricing. 
This may provide new ideas and inspiration for future research in the field.

Nonetheless, several issues remain to be resolved, such as the overfitting tendency of the KANs model and the potential effects of replacing the fully connected layer of Conv-KANs with KANs. 
These questions are left for future investigators to address.

Overall, the addition of KANs has made the field of options pricing more dynamic. Moreover, convolutional versions of neural networks may be one of the answers to option pricing.

\section{Declaration of competing interest}
The authors declare that they have no known competing financial 
interests or personal relationships that could have appeared to influence 

\section{Declaration of generative AI and AI-assisted technologies in the writing process}
During the preparation of this work the author(s) used ChatGPT 4o in order to polish the sentences. After using this tool/service, 
the author(s) reviewed and edited the content as needed and take(s) full responsibility for the content of the published article.

\appendix
\section{Stochastic process of no-dividend-paying stock}
\label{app1}
The stochastic process assumed for the price of a no-dividend-paying stock is examined. It is important to highlight, prior to beginning, that when analyzing stocks or other financial products, 
it is crucial to focus on ratios rather than absolute values. For example, if two different derivatives are invested in financial markets and a gain of 50 cents is made, but the amounts invested are not the same—one is \$1, the other is \$2—the resulting yields will differ.

When it comes to the drift rate of stocks, it should be assumed that the form of the drift rate is related to the price of the stock: 
If $S$ is the stock price at time $t$, and the mean drift rate related to $S$ should be $uS$ for certain constant parameter $\mu $. 
And This means that in a short interval of time $\Delta t$, the expected increase in $S$ is $S\Delta t$. The parameter $\mu $ is the expected rate of return on the stock [15]. 
If the coefficient of $dz$ is zero, then this model implies that
\begin{equation}
  \Delta S=\mu S\Delta t
\end{equation}
in the limit, as $\Delta t\rightarrow 0$, so that
\begin{equation}
  dS=\mu Sdt
\end{equation}
After integrating, this gives
\begin{equation}
  S_T=S_0e^{\mu T}
\end{equation}
Where $S_0$ and $S_T$ respectively represent the price of stock at time 0 and $T$. 
On this basis, the parameter $\sigma $ is introduced to represent the volatility, and we have
\begin{equation}
  dS=\mu Sdt+\sigma Sdz
\end{equation}
or

\centerline{$\frac{dS}S=\mu dt+\sigma dz$}

Equation (A.4) is the most widely used model of stock price behavior. More information about $dz$ see paper [16].

\section{Derivation of the Black-Scholes-Merton differential equation}
The price of derivatives at time $t$ is analyzed. If $T$ represents the maturity date, the time to maturity is given by $T-t$.

Formula (A.4) is used to characterize the underlying stock in section Appendix A. That is

\centerline{$dS=\mu Sdt+\sigma Sdz$}

Suppose $f$ is the call option contingent $S$. The variable $f$ must be function of $S$ and $t$, that is $f(S, t)$. From [17] we obtain Itô lemma
\begin{equation}
  dG=\left(\frac{\partial G}{\partial x}a+\frac{\partial G}{\partial t}+\frac{1}{2}\frac{\partial^2G}{\partial x^2}b^2\right)dt+\frac{\partial G}{\partial x}bdz
\end{equation}
where

\centerline{$dx=a(x,t)dt+b(x,t)dz$}

$dx$ is Itô process and $G$ is a function of $x$ and $t$, $dz$ is wiener process. Itô lemma. Then substitute (A.4) into (B.1), and obtain
\begin{equation}
  dG=\left(\frac{\partial G}{\partial x}\mu S+\frac{\partial G}{\partial t}+\frac{1}{2}\frac{\partial^2G}{\partial x^2}\sigma^2S^2\right)dt+\frac{\partial G}{\partial x}\sigma Sdz
\end{equation}
Substitute $f$ for $G$, we have
\begin{equation}
  df=\left(\frac{\partial f}{\partial S}\mu S+\frac{\partial f}{\partial t}+\frac12\frac{\partial^2f}{\partial S^2}\sigma^2S^2\right)dt+\frac{\partial f}{\partial S}\sigma Sdz
\end{equation}
The discrete versions of equations (A.4) and (B.3) are respectively
\begin{equation}
  \Delta S=\mu S\Delta t+\sigma S\Delta z
\end{equation}
and
\begin{equation}
  \Delta f=\left(\frac{\partial f}{\partial S}\mu S+\frac{\partial f}{\partial t}+\frac{1}{2}\frac{\partial^2f}{\partial S^2}\sigma^2S^2\right)\Delta t+\frac{\partial f}{\partial S}\sigma S\Delta z
\end{equation}

Equations (B.4) and (B.5) have the same $\Delta z$. An appropriate asset portfolio $P$ can be chosen to eliminate $\Delta z$. That is

\begin{center}
  -1: derivative \\
  $+\partial f/\partial S$: stocks 
\end{center}

The holder of this portfolio is short one derivative and long an amount $\partial f/\partial S$ of stocks. Give the definition as the value of the portfolio by 
\begin{equation}
  P=-f+\frac{\partial f}{\partial S}S
\end{equation}
and its discrete form
\begin{equation}
  \Delta P=-\Delta f+\frac{\partial f}{\partial S}\Delta S
\end{equation}
Substituting equations (B.4) and (B.5) into equation (B.7) yields
\begin{equation}
  \Delta P=\left(-\frac{\partial f}{\partial t}-\frac{1}{2}\frac{\partial^2f}{\partial S^2}\sigma^2S^2\right)\Delta t
\end{equation}

Clearly, equation (B.8) does not have relation to $\Delta z$. Prove that in time $\Delta t$, the portfolio maintains risk-free feature. 
In the view of this property, It is proposed to use the risk-free rate $f$ at a given time in the financial markets to replicate the portfolio through its yields.
\begin{equation}
  \Delta P=rP\Delta t
\end{equation}
Through substitution, calculation and integration, this implies that
\begin{equation}
  \frac{\partial f}{\partial t}+rS\frac{\partial f}{\partial S}+\frac12\sigma^2S^2\frac{\partial^2f}{\partial S^2}=rf
\end{equation}

Equation (B.10) is the Black-Scholes-Merton differential equation.



\begin{thebibliography}{00}

\bibitem[Black F, Scholes M(1973)]{B-S73}
Black F, Scholes M. The pricing of options and corporate liabilities[J]. 
Journal of political economy, 1973, 81(3): 637-654.
\bibitem[Merton R C(1973)]{Merton73}
Merton R C. Theory of rational option pricing[J]. 
The Bell Journal of economics and management science, 1973: 141-183.
\bibitem[Ge M, Zhou S, Luo S, et al.(2021)]{Ge M21}
Ge M, Zhou S, Luo S, et al. 3D Tensor-based Deep Learning Models for Predicting Option Price[C].
2021 International Conference on Information Science and Communications Technologies (ICISCT). IEEE, 2021: 1-6.
\bibitem[Ivașcu C F(2021)]{Ivașcu C F21}
Ivașcu C F. Option pricing using machine learning[J]. 
Expert Systems with Applications, 2021, 163: 113799.
\bibitem[Kodratoff Y, Manago M, Blythe J(1987)]{Kodratoff Y87}
Kodratoff Y, Manago M, Blythe J. Generalization and noise[J]. 
International Journal of Man-Machine Studies, 1987, 27(2): 181-204.
\bibitem[Shi X, Chen Z, Wang H, et al.(2015)]{Shi X15}
Shi X, Chen Z, Wang H, et al. Convolutional LSTM network: A machine learning approach for precipitation nowcasting[J]. 
Advances in neural information processing systems, 2015, 28.
\bibitem[Liu Z, Wang Y, Vaidya S, et al.(2024)]{Liu Z24}
Liu Z, Wang Y, Vaidya S, et al. Kan: Kolmogorov-arnold networks[J]. 
arXiv preprint arXiv:2404.19756, 2024.
\bibitem[Bodner A D, Tepsich A S, Spolski J N, et al.(2024)]{Bodner A D24}
Bodner A D, Tepsich A S, Spolski J N, et al. Convolutional Kolmogorov-Arnold Networks[J]. 
arXiv preprint arXiv:2406.13155, 2024.
\bibitem[Drokin I(2024)]{Drokin I24}
Drokin I. Kolmogorov-Arnold Convolutions: Design Principles and Empirical Studies[J]. 
arXiv preprint arXiv:2407.01092, 2024.
\bibitem[Hutchinson J M, Lo A W, Poggio T(1994)]{Hutchinson J M94}
Hutchinson J M, Lo A W, Poggio T. A nonparametric approach to pricing and hedging derivative securities via learning networks[J]. 
The journal of Finance, 1994, 49(3): 851-889.
\bibitem[Garcia R, Gençay R(2000)]{Garcia R, Gençay R00}
Garcia R, Gençay R. Pricing and hedging derivative securities with neural networks and a homogeneity hint[J]. 
Journal of Econometrics, 2000, 94(1-2): 93-115.
\bibitem[Gençay R, Qi M(2001)]{Gençay R01}
Gençay R, Qi M. Pricing and hedging derivative securities with neural networks: Bayesian regularization, early stopping, and bagging[J]. 
IEEE transactions on neural networks, 2001, 12(4): 726-734.
\bibitem[Arifovic J, Gencay R(2001)]{Arifovic J01}
Arifovic J, Gencay R. Using genetic algorithms to select architecture of a feedforward artificial neural network[J]. 
Physica A: Statistical mechanics and its applications, 2001, 289(3-4): 574-594.
\bibitem[Zouaoui H, Naas M N(2023)]{Zouaoui H23}
Zouaoui H, Naas M N. Option pricing using deep learning approach based on LSTM-GRU neural networks: Case of London stock exchange[J]. 
Data Science in Finance and Economics, 2023, 3(3): 267-284.
\bibitem[Hull J C, Basu S (2016)]{Hull J C16}
Hull J C, Basu S. Options, futures, and other derivatives[M]. 
Pearson Education India, 2016.
\bibitem[Major P(1981)]{Major P81}
Major P, Major P. Multiple Wiener-Itô integrals[M]. 
Springer Berlin Heidelberg, 1981.
\bibitem[Itô K(1951)]{Itô51}
Itô K. On stochastic differential equations[M]. 
American Mathematical Soc., 1951
\bibitem[DeVore R A(1993)]{DeVore93}
DeVore R A, Kyriazis G, Leviatan D, et al. Wavelet compression and nonlinear n-widths[J]. 
Adv. Comput. Math., 1993, 1(2): 197-214.
\bibitem[Koenig B C(2024)]{Koenig24}
Koenig B C, Kim S, Deng S. KAN-ODEs: Kolmogorov-Arnold network ordinary differential equations for learning dynamical systems and hidden physics[J].
Computer Methods in Applied Mechanics and Engineering, 2024, 432: 117397.
\bibitem[Li Y(2018)]{Li18}
Li Y, Jiang W, Yang L, et al. On neural networks and learning systems for business computing[J]. 
Neurocomputing, 2018, 275: 1150-1159.
\bibitem[Sun B Q(2015)]{Sun15}
Sun B Q, Guo H, Karimi H R, et al. Prediction of stock index futures prices based on fuzzy sets and multivariate fuzzy time series[J].
Neurocomputing, 2015, 151: 1528-1536.
\bibitem[Ma C(2021)]{Ma21}
Ma C, Zhang J, Liu J, et al. A parallel multi-module deep reinforcement learning algorithm for stock trading[J]. 
Neurocomputing, 2021, 449: 290-302.
\bibitem[Yang S(2024)]{Yang24}
Yang S, Tang D. A large-scale microblog dataset and stock movement prediction based on Supervised Contrastive Learning model[J].
Neurocomputing, 2024, 584: 127583.
\bibitem[Xu H(2020)]{Xu20}
Xu H, Chai L, Luo Z, et al. Stock movement predictive network via incorporative attention mechanisms based on tweet and historical prices[J]. 
Neurocomputing, 2020, 418: 326-339.
\end{thebibliography}



\end{document}